\documentclass[12pt]{article}
\usepackage{amssymb}

\title{Feynman's Path Integrals and Bohm's Particle Paths}
\author{
Roderich Tumulka\footnote{Dipartimento di Fisica dell'Universit\`a
     di Genova and INFN sezione di Genova, Via Dodecaneso 33, 16146
     Genova, Italy. E-mail: tumulka@mathematik.uni-muenchen.de}
}
\date{February 18, 2005} 

\newcommand{\RRR}{\mathbb{R}}
\newcommand{\CCC}{\mathbb{C}}
\newcommand{\vQ}{Q}

\newcommand{\im}{\mathrm{Im}}

\newcommand{\Q}{\mathcal{Q}}

\begin{document}
\maketitle

\begin{abstract}
  Both Bohmian mechanics, a version of quantum mechanics with
  trajectories, and Feynman's path integral formalism have something
  to do with particle paths in space and time. The question thus
  arises how the two ideas relate to each other. In short, the answer
  is, path integrals provide a re-formulation of Schr\"odinger's
  equation, which is half of the defining equations of Bohmian
  mechanics.  I try to give a clear and concise description of the
  various aspects of the situation.

\medskip

\end{abstract}

\section{Background}

Bohmian mechanics \cite{Bohm52, Hol93, survey, Gol01, Tum04}, invented
by D.~Bohm in 1952, is a (nonrelativistic) quantum theory without
observers; in a world governed by this theory, quantum particles have
precise trajectories, and observers find the statistics of outcomes of
their experiments in agreement with quantum mechanics.  For a system
of $N$ particles, their positions $\vQ_i(t) \in \RRR^3$ change
according to the equation of motion,
\begin{equation}\label{Bohm}
  \frac{d\vQ_i(t)}{dt} = \frac{\hbar}{m_i} \im \frac{\psi_t^* \nabla_i
  \psi_t}{\psi_t^* \psi_t} \bigl( \vQ_1(t), \ldots, \vQ_N(t) \bigr)
  \,.
\end{equation}
Here, $m_i$ is the mass of particle $i$, $\phi^*\psi$ denotes the
scalar product in $\CCC^k$, and $\psi_t: \RRR^{3N} \to \CCC^k$ is the
wave function of non-relativistic quantum mechanics, defined on the
configuration space and evolving according to the (nonrelativistic)
Schr\"odinger equation.
In a typical Bohmian universe, the positions $\vQ_i(t)$ appear random,
at any time $t$, with joint distribution $|\psi_t|^2$ \cite{survey,
Gol01}.

The method of path integrals \cite{Fey48, FH65, Sch81}, invented by
R.~P.~Feynman in 1942, is nowadays widely used in quantum physics.  I
briefly recall the key idea.  Let us assume for simplicity a finite
set $\Q$ as configuration space (just as a mathematical model), and
that time is discrete, too. Then the time evolution of the wave
function $\psi$ is defined by a unitary operator $U$ representing one
time step,
\begin{equation}
  \psi_{t+1}(r) = \sum_{q \in \Q} U(r,q) \, \psi_t(q) \,,
\end{equation}
where $q$ and $r$ run through all configurations, and $U(r,q)$ are the
matrix elements of $U$ in the position representation. It follows that
after $s$ time steps,
\begin{equation}
  \psi_{t+s}(r) = \sum_{q_t, \ldots, q_{t+s-1} \in \Q} U (r,q_{t+s-1})
  \cdots U(q_{t+2},q_{t+1}) \, U (q_{t+1},q_{t} ) \, \psi_t (q_t) \,.
\end{equation}
Since the sum is over all paths, what we have is a path integral!
What I want to point out with this little calculation is that path
integrals are a mathematical formulation of the \emph{time evolution}
of $\psi$, an equivalent alternative to the Schr\"odinger equation.
The more general formula, for continuous time and configuration space
$\Q$, reads
\begin{equation}
  \psi_{t+s}(r) = \int D\gamma \, e^{iS[\gamma]/\hbar} \,
  \psi_t(\gamma_t) \,,
\end{equation}
where the integration is over all paths $\gamma$ in $\Q$ during the
time interval $[t,t+s]$ with $\gamma_{t+s} = r$, and $S[\gamma]$
denotes the (classical) action of the path $\gamma$.

\section{Discussion}

Since the same $\psi$ as in usual quantum mechanics appears in Bohmian
mechanics, and with the same time evolution, path integrals are
equally relevant to Bohmian mechanics as to usual quantum mechanics. 
Whatever can be calculated about the evolution by means of path
integrals in ordinary quantum mechanics is equally correct in Bohmian
mechanics. Note that the path integrals concern \emph{only} the wave
function, not the Bohmian paths.  It may be helpful at this point to
keep in mind that path integrals are not exclusive to quantum theory:
one can just as well re-write Maxwell's equations of classical
electrodynamics, as in fact any linear field equation, in terms of
path integrals. 

An obvious but basic fact that I want to emphasize is that the paths
of the path integral have a very different status from the Bohmian
paths: \textit{Feynman's paths are mathematical tools for computing
the evolution of $\psi$, while one among Bohm's paths is the actual
motion of the Bohmian particle, which exists in addition to $\psi$.} 

Now there is another, completely different relation between path
integrals and theories of actual particle trajectories: since one
considers a lot of possible paths of the particle, is it perhaps one
of these paths that the particle actually follows? If the path
integral formalism provided a probability distribution on the space of
all paths, one could assume that nature chooses one of the paths at
random according to this distribution. This theory would have the same
ontology as Bohmian mechanics, but a different, in fact stochastic,
law of motion. That would seem like a completely reasonable and
acceptable theory, and if it predicted correctly all probabilities for
measurements, as prescribed by the quantum formalism, it would be a
serious candidate for the explanation of how the quantum world really
works. 

But do we have a probability measure on path space? One of the central
objects in the path integral formalism is a \emph{complex} measure
$D\gamma \, e^{iS[\gamma]/\hbar}$ on path space (though it is not
rigorously defined in the continuum case, see the last paragraph
below). To obtain a real measure, we could take, for example, its
positive real or its positive imaginary part or its modulus, but the
probabilities for the outcomes of measurements entailed by that
measure will not agree with quantum mechanics (in contrast to those in
Bohmian mechanics).  A better probability measure on path space is not
in sight.

Thus, the formulation ``possible paths of the particle'' for the paths
considered in a path integral, a formulation that comes to mind rather
naturally, \emph{cannot}, in fact, be taken literally.  The status of
the paths is more like ``possible paths along which a part of the
\emph{wave} may travel,'' to the extent that waves travel along paths. 
For example, in the double-slit experiment some paths pass through one
slit, some through the other; correspondingly, part of the wave passes
through one slit and part through the other. 

(In brackets, a few more remarks on measures on path space.  When one
inserts imaginary numbers for the time variable, the complex measure
$D\gamma \, e^{iS[\gamma]/\hbar}$ on path space turns into a
probability measure defining a stochastic process \cite{Sch81}; the
corresponding path integrals are known as Feynman--Kac integrals and
useful for computing the operators $e^{-\beta H}$ instead of
$e^{-(i/\hbar)tH}$. But the process does not, of course, provide us
with a quantum theory without observers, as physical time is not
imaginary.)

Finally, I address mathematical rigor.  There is a mathematical
problem with path integrals when both the time axis and configuration
space are taken to be a continuum, a problem that is absent when
either one is taken to be a discrete set.  The problem is that there
exists no analogue to Lebesgue measure, in other words no natural
notion of volume, on the space of all paths, because this space is
infinite-dimensional.  Such a measure $D\gamma$, however, is assumed
in the path integral formalism.  This problem is not cured by Bohmian
mechanics, and, as far as I can see, the problem will never be cured
because it is simply a mathematical fact that there is no Lebesgue
measure on infinite-dimensional spaces. (However, many astute
proposals have been made as to how path integrals can be defined
rigorously if one does not insist that it actually be the integral of
the action functional $e^{iS[\gamma]/\hbar}$ over path space
\cite{AHK76}.)  I emphasize, however, that path integrals, even when
mathematically not clean, often provide correct results by means of a
comparatively simple calculation.

\section{Acknowledgments}

This note was inspired by questions from Dennis Smoot of the
University of Illinois, Chicago, USA. I thank Jean Bricmont of
Universit\'e catholique de Louvain, Belgium, for his helpful comments.
This work was supported by INFN.


\begin{thebibliography}{10}

\bibitem{Bohm52} 
D.~Bohm. 
A Suggested Interpretation of the Quantum
   Theory in Terms of ``Hidden'' Variables, I and II. 
{\em Phys. Rev.}, 85:166--193, 1952. 

\bibitem{Hol93} 
P.~R.~Holland. 
{\em The Quantum Theory of Motion.} 
Cambridge University Press, 1993. 

\bibitem{survey} 
K.~Berndl, M.~Daumer, D.~D\"urr, S.~Goldstein, and N.~Zangh\`\i. 
A survey of Bohmian mechanics. 
{\em Il Nuovo Cimento}, 110B:737--750, 1995. 

\bibitem{Gol01}
S.~Goldstein. 
Bohmian mechanics. 2001. 
In {\em Stanford Encyclopedia of Philosophy}. Ed.~by
 E.~N.~Zalta, published online by Stanford University. \\
http://plato.stanford.edu/entries/qm-bohm/

\bibitem{Tum04}
R.~Tumulka. 
Understanding Bohmian mechanics: A dialogue. 
{\em Am. J. Phys.}, 72(9):1220--1226, 2004. 

\bibitem{Fey48} 
R.~P.~Feynman.  
Space-time approach to non-relativistic quantum mechanics.  
{\em Rev. Mod. Phys.}, 20:367--387, 1948. 

\bibitem{FH65}
R.~P.~Feynman and A.~R.~Hibbs. 
{\em Quantum Mechanics and Path Integrals.} 
McGraw-Hill, New York, 1965. 

\bibitem{Sch81}
L.~S.~Schulman. 
{\em Techniques and Applications of Path Integration}. 
John Wiley \& sons, New York, 1981. 

%
%

\bibitem{AHK76}
S.~A.~Albeverio and R.~J.~H{\o}egh-Krohn. 
{\em Mathematical theory of Feynman path integrals}. 
 Lecture Notes in Mathematics, Vol. 523. 
Springer-Verlag, Berlin-New York, 1976. 



\end{thebibliography}
\end{document}